\documentclass[12pt]{article}

\pdfoutput=1

\usepackage[usenames,dvips]{color}
\usepackage{amssymb}
\usepackage{amsmath}
\usepackage{epsfig}
\usepackage[utf8]{inputenc}
\usepackage[T1]{fontenc}
\usepackage{graphicx}
\usepackage{xcolor}
\usepackage{tikz}

\newcommand{\MeV}{{\rm MeV}}

\newcommand{\GeV}{{\rm GeV}}

\newcommand{\Mp}{M_{\rm P}}
\def\d{{\rm d}}

\begin{document}
%%%%%%%%%%%%%%%%%%%%%%%%%%%%%%%%%%%%%%%%%%%%%%%%%%%%%%%%%%%%%%

\thispagestyle{empty}

\title{\bf\boldmath Testing $\alpha$-attractor P-model of inflation\\[4pt] 
by Cosmic Microwave Background radiation\thanks{Dedicated to Andrzej Białas for his 90th Birthday. To be published in Acta Physica Polonica B}
}
\author{Michał Marciniak, Marek Olechowski, Stefan Pokorski\\[6pt]
\em
Institute of Theoretical Physics, Faculty of Physics, University of Warsaw\\[2pt]
\em
Pasteura 5, 02-093 Warsaw, Poland}
\date{}
\maketitle
\begin{abstract}
%\centerline{Dedicated to Andrzej Białas for his 90th Birthday}
%\medskip

In a recently proposed approach to testing models of inflation  by Cosmic Microwave Background (CMB) radiation the reheating temperature is directly expressed in terms of the CMB observables.
Its model independent bounds translate  in a given model into narrow ranges of those observables.  In  that approach we analyse the polynomial class of the $\alpha$-attractor inflaton potential models (P-models), in a broad range of polynomials  and with the inflaton decays and fragmentation in the reheating period taken into account. The predictions for the CMB observables, the scalar spectral index $n_s$ and tensor-to-scalar ratio $r$, are compared with the Planck and Planck combined with ACT data. Both can be accommodated by that class of the $\alpha$ attractor models. The sensitivity of the results of that comparison to the reheating temperature and to the upper bound on the ratio $r$ is clearly demonstrated.
\end{abstract}
  
\section{Introduction}

The inflation paradigm solves beautifully the large scale homogeneity and flatness problem of the universe.  The thermal history of the universe,  beginning with the  primordial big bang nucleosynthesis (BBN),  is well understood in terms of the Standard Model (SM) of elementary interactions.    
However,  very little is known about the cosmological history from after the end of inflation until the BBN, which may span even more than 30 orders of magnitude in time scales.  And it is in that period when the two main cosmological puzzles,  production of dark matter (DM) and generation of baryon asymmetry (BA),  must have happened.  Moreover,  it is well known that some physics beyond the SM is necessary for understanding those puzzles.

According to the successful thermal history, during the BBN era  the universe was dominated by radiation (RD). This implies that there must have been a period, called reheating of the universe, in which the empty and cold universe after inflation evolved into the hot universe of RD era.  Actually,  the RD era could have started long before BBN.  According to the inflationary paradigm, in that process the energy stored in the inflaton field was used to produce plasma of relativistic particles. 
A crucial parameter linked to the RD era is the reheating temperature $T_{\rm re}$,  the temperature of the  plasma at the beginning of the RD phase. The reheating  temperature depends on the details  of the  inflation process, behaviour of the inflaton potential after the end of inflation and  on the reheating mechanism. 
It provides the most important link  between the evolution of the early universe and beyond the SM physics, necessary to understand the DM production and the BA generation, that 
could had happened
either during the reheating period or already in the RD and was sensitive to the reheating temperature.

It has been emphasized for a long time that the measurement of the CMB radiation can give us an important insight into the inflationary and reheating periods,  constraining theoretical models of both.  
The results from WMAP \cite{WMAP:2003elm,WMAP:2012fli},  Planck \cite{Planck:2018vyg,Planck:2018jri}  and BICEP/Keck \cite{BICEP2:2018kqh,BICEP:2021xfz} for the spectral index $n_s$ of the power spectrum of the scalar perturbations and the upper bound  for the  ratio of the tensor to scalar perturbations $r$ 
have provided  quantitative constraints on the theoretical models. More recently, there have been released the results of the  ground-bases experiment,  Atacama Cosmology Telescope (ACT) \cite{AtacamaCosmologyTelescope:2025blo,AtacamaCosmologyTelescope:2025nti}.
The combination of the Planck + BICEP/Keck + ACT data in conjunction with the results of the Baryon Acoustic Oscillations (BAO) measured by the Dark Energy Spectroscopic Instrument (DESI) \cite{DESI:2024uvr,DESI:2024mwx} shift $n_s$ towards larger values \cite{AtacamaCosmologyTelescope:2025nti} than  the Planck + BICEP/Keck + DESI (P-BK-D) results \cite{BICEP:2021xfz} (for the  extensive discussion of the experimental situation see \cite{Ellis:2025zrf}).
The new data and the global fit to the Planck + BICEP/Keck + DESI + ACT (P-BK-D-ACT) results have further invigorated   
theoretical research on inflationary and reheating models 
\cite{Kallosh:2025rni,Aoki:2025wld,Berera:2025vsu,Brahma:2025dio,Dioguardi:2025vci,Gialamas:2025kef,Salvio:2025izr,Kim:2025dyi,Antoniadis:2025pfa,Dioguardi:2025mpp,Gao:2025onc,He:2025bli,Drees:2025ngb,Zharov:2025zjg,Haque:2025uri,Liu:2025qca,Yin:2025rrs,Gialamas:2025ofz,Haque:2025uis,Yogesh:2025wak,Byrnes:2025kit,Yi:2025dms,Addazi:2025qra,Maity:2025czp,Peng:2025bws,Mondal:2025kur,Kallosh:2025ijd,Haque:2025uga,Pallis:2025nrv,Choudhury:2025vso,Odintsov:2025wai,Wolf:2025ecy}. 
It has been shown that the details of the inflaton potentials,  modifying their approximate plateau,  change their predictions  and make them consistent with one or another set of data.  The expected improvement in the precision of the cosmological measurements will of course further constrain the models.

In a recent paper \cite{Ghoshal:2025ejg} there have been investigated  several classes of the $\alpha$-attractor models of inflation \cite{Kallosh:2013lkr,Kallosh:2013hoa,Kallosh:2013yoa,Kallosh:2013pby,Kallosh:2013maa,Galante:2014ifa,Kallosh:2022feu}, the  so called  E-, T- and P-models.
A systematic framework is presented that expresses all free parameters of a given inflationary potential directly in terms of CMB observables $(A_s, n_s, r)$, and then uses an one-parameter reheating model (characterised by an inflaton equation-of-state parameter $w$ and a dissipation rate $\Gamma$, or equivalently a reheating temperature $T_{\rm re}$) to derive a consistency relation that uniquely fixes $T_{\rm re}$ in terms of the same observables. 
In ref.~\cite{Ghoshal:2025ejg} the emphasize is on the direct link of the CMB observables to the reheating temperature $T_{\rm re}$ instead of the number of e-folds $N_k$ between the (event) horizon exit of the experimentally observed perturbation with a comoving wave number $k$ and the end of inflation.  That approach is useful because there are model independent bounds on $T_{\rm re}$ that can be used to constrain models of inflation  and the reheating dynamics.  And,  vice versa,  with an improvement in the precision of the CMB data one can get stronger bounds on $T_{\rm re}$.  It is shown in ref.~\cite{Ghoshal:2025ejg} that  the model independent bounds on the $T_{\rm re}$ give different for different models of inflation very narrow ranges of the power spectrum index $n_s$,  and with  interesting dependence on the value of $r$.  A comparison of those predictions with the Planck,  ACT and DESI data has been presented.  For instance,  it has been shown that  the Starobinsky model of inflation (which is the special case of the E-$\alpha$-attractor model  is excluded at the $2\sigma$ level by the model independent bounds on the $T_{\rm re}$ and the  
global fit  to  $n_s$  given by the combination of the P-BK-D-ACT results whereas it remains consistent with  the P-BK-D combination.  More general versions of the three classes of models remain consistent with both combinations of the experimental results, however with some model dependent constraints on the acceptable range of the $T_{\rm re}$

In the present paper we apply similar approach as in ref.~\cite{Ghoshal:2025ejg} to a more detailed studies of the polynomial P $\alpha$-attractor models, which are consistent with the results of the P-BK-D combination in a broader range of model parameters and of the $T_{\rm re}$. Furthermore, we discuss the impact of some non-perturbative effects in the reheating dynamics.

%%%%%%%%%%%%%%%%%%%%%%%%%%%%%%%%%%%%%%%%%%
\section{A brief recollection of the approach in ref.~\cite{Ghoshal:2025ejg}}
%%%%%%%%%%%%%%%%%%%%%%%%%%%%%%%%%%%%%%%%%%

At present, the  CMB  data give us the value of the spectral index $n_s(k_*)$
for the scalar perturbation mode with the (pivot) wavenumber $k_*/a_0 =0.05\,$Mpc$^{-1}$, its amplitude  $A_s(k_*)$
\cite{Planck:2018vyg} and the upper bound on the ratio of the tensor to scalar perturbation amplitudes $r(k_*)$. Following the inflation paradigm, the values of those observables depend on the shape of the inflaton potential, its  independent parameters and the value of the inflaton field $\phi_{k_*}$ when the mode $k_*$ exits the horizon during inflation.

In the slow-roll approximation the observables $n_s(k)$, $r(k)$, and $A_s(k)$, where $k$ is some comoving wavenumber of the metric perturbations, are related to the inflaton potential as follows:
\begin{eqnarray}
        n_s(k) &=& 1 - 6\epsilon_k + 2 \eta_k\,,
        \label{eq:ns(k)}
        \\
        r(k) &=& 16\epsilon_k\,,
        \label{eq:r(k)} 
        \\
        A_s(k) &=& \frac{V(\phi_k)}{24 \pi^2 \epsilon_k \Mp^4 }\,,
        \label{eq:As(k)}
\end{eqnarray}
where $\Mp$ is the reduced Planck mass 
and $\phi_k$ is the value of the inflation field at time when the perturbation with comoving wavenumber $k$ was generated.
The parameters $\epsilon_k$ and $\eta_k$  read:
\begin{eqnarray}
        \epsilon_k&=&\left.\frac{1}{2}\Mp^2 \left(\frac{\partial_\phi V(\phi)}{V(\phi)}\right)^2\right|_{\phi=\phi_k}\;, 
       \label{eq:epsilon_k}
        \\
        \eta_k &=& \left.\Mp^2\,\frac{\partial^{(2)}_\phi V({\phi})}{V(\phi)}\right|_{\phi=\phi_k}\;,
        \label{eq:eta_k}
\end{eqnarray}

We focus now on  one  classes of the $\alpha$-attractor inflaton models, the so-called  P-models,  which can better describe  the combined PACT data \cite{AtacamaCosmologyTelescope:2025nti}.  Its potential reads
\cite{Kallosh:2013lkr,Kallosh:2013hoa,Kallosh:2013yoa,Kallosh:2013pby,Kallosh:2013maa,Galante:2014ifa,Kallosh:2022feu}: 
\begin{equation}
 V(\phi,\alpha,n,\Lambda_{inf})=\Lambda_{\mathrm{\rm inf}}^4\frac{\phi^{2n}}{\phi^{2n}+\left( \sqrt{\frac{3\alpha}{2}}\Mp\right)^{2n}}
    \label{eq:V}
\end{equation}  \\
where $\Lambda_{\rm inf}$ represents a mass scale that determines the energy scale of the inflation and $\sqrt{3\alpha/2}\Mp$ is an effective scale that can be higher than $\Mp$.\

All the perturbation parameters $n_s(k)$ etc., expressed in terms of the potential parameters, read (see eqs.~\eqref{eq:ns(k)}-\eqref{eq:V}):
\begin{eqnarray}
A_s
&=&
\frac{1}{48\pi^2n^2}\frac{\Lambda_{\rm inf}^4\phi_k^2}{\Mp^6}
\left(\frac{2\phi_k^2}{3\alpha \Mp^2}\right)^{n}
\left(1+\left(\frac{2\phi_k^2}{3\alpha \Mp^2}\right)^{n}\right)
\,,\label{eq:As}
\\
n_s
&=&
1-4n\frac{\Mp^2}{\phi_k^2}\,
\left(n+1+(2n+1)\left(\frac{2\phi_k^2}{3\alpha \Mp^2}\right)^{n}\right)
\left(1+\left(\frac{2\phi_k^2}{3\alpha \Mp^2}\right)^{n}\right)^{-2}
\!\!\!,\qquad\label{eq:ns}
\\
r
&=&
32n^2\frac{\Mp^2}{\phi_k^2}\,
\left(1+\left(\frac{2\phi_k^2}{3\alpha \Mp^2}\right)^{n}\right)^{-2}.
\label{eq:r}
\end{eqnarray}

The parameter $n$ in the  exponent  in the potential \eqref{eq:V} can be an integer or it can take fractional values and we consider it as a number which defines the model.\footnote{
We thank Renata Kallosh and Andrei Linde for drawing our attention to the fact that for polynomial potentials like P-model fractional values of $n$ are also of theoretical interest.}
The other parameters $\Lambda_{\rm inf}, \alpha$  and the inflaton field value  $\phi_k$ when the mode with the co-moving wavenumber $k$ left the horizon can be expressed  in terms of the three observables by the inverse relations.
It is convenient to define the following combination of the CMB observables,  $n_s$ and $r$, and the  potential parameter $n$:
\begin{equation}
\label{eq:xi}
\xi=n\left(8\,\frac{1-n_s}{r}-1\right)\,.
\end{equation}
We get then:
\begin{eqnarray}
\alpha
&=&
\frac{64n^2}{3r}\left(\frac{2n+1}{2n+\xi}\right)^2\sqrt[n]{\frac{2n+1}{\xi-1}}
\label{eq:alpha}
\,,\\
\phi_k
&=&
\Mp\sqrt{\frac{32}{r}}\,n\,\frac{2n+1}{2n+\xi}
\label{eq:phi_k}
\,,\\
\Lambda_{\rm inf}^4
&=&
\frac{3\pi^2}{2}\,\Mp^4\,rA_s\left(\frac{\xi-1}{2n+\xi}\right)\,.
\label{eq:Lambda4}
\end{eqnarray}
Thus, the values of the three observables for some comoving wavenumber $k$ fully determine the parameters of the $\alpha$-attractor potential for a given choice of the exponent $2n$.

The experimental results are used to determine CMB observables and usually presented as the experimentally allowed regions (at some confidence level) in the plane $(n_s(k_*), r(k_*))$ where $k_*$ is a conveniently chosen pivot scale (in our calculations we use  $k=k_* = a_0\times 0.05$\,Mpc$^{-1}$).
It is customary to compare them with models of inflation by checking if those allowed regions are consistent with the \textbf{assumed} number of e-folds, usually $50\div60$, during the rolling down of the inflaton field from its value $\phi_{k_*}$ to the value $\phi_{\rm end}$ at the end of inflation. 
However, the a priori acceptable range of $N_k$ can actually be very large and is correlated with the expansion during the reheating period (see e.g.~textbooks \cite{Kolb:1990vq,Weinberg:2008zzc}. It is then interesting, instead of assuming a value (or range of values) of $N_k$, to  calculate it, given some assumptions about the reheating period.

The value of $\phi_{\rm end}$ at the end of (slow-roll) inflation can be estimated by the conditions for the slow-roll parameters $\epsilon =1$ or $|\eta |=1$, whichever is reached earlier.  Contrary to E- and T- models,  it is not possible to get closed expression for those conditions and solve them for $\phi_{\rm end}^{(\epsilon)}$ in P-models 
for arbitrary value of the parameter $n$ because it is related to the solution of equation $x(1+x^{2n})={\rm const}$. One obvious exception is $n=\frac12$. The solutions in such a case reads
\begin{equation}
\phi_{\rm end}^{(n=1/2)}
=
2\sqrt{2}\frac{\sqrt{4+\sqrt{r}\left(\xi^2-1\right)}-2}{\sqrt{r}\left(\xi^2-1\right)}\,M_P\,.
\label{eq:phi_end_1/2}
\end{equation}
The explicit solutions in the case of $n=1$ can also be found but they are rather lengthy and complicated. For other values of $n$ solutions for $\phi_{\rm end}$ can be easily found numerically.

The number of e-folds, $N_k$, is expressible via $\phi_k$ and $\phi_{\rm end}$: 
\begin{equation}
   N_k =-\frac{1}{\Mp^2}\int_{\phi_k}^{\,\phi_{\rm end}}\frac{V(\phi)}{\partial_\phi V(\phi)}\,{\rm d}\phi\,.
   \label{eq:Nk-integral}
\end{equation}
Thus, using eqs.~\eqref{eq:phi_k} and expressions (or numerical results) for $\phi_{\rm end}$, one finds that $N_k$ is determined by the observables $n_s$ and $r$. For P-model with $n=\frac12$ eq.~\eqref{eq:phi_end_1/2} may be used to obtain
\begin{equation}
N_k^{(n=1/2)}
=
\frac{2\left(2\xi-\sqrt{4+\sqrt{r}\left(\xi^2-1\right)}\right)}{\sqrt{r}\left(\xi^2-1\right)}\,.
\label{eq:Nk_1/2}
\end{equation}

For each considered model the number of e-folds $N_{k_*}$ may be calculated for each pair of values of the CMB observables $n_s(k_*)$ and $r(k_*)$. However, not every pair of such values is consistent with the model of inflation
used for calculating $N_{k_*}$ and the assumed description of the reheating
process. One of the reasons is related to the model independent bounds on the reheating temperature which characterizes the transition from the reheating era to the radiation domination era. In our approach this temperature is a function of the CMB observables and the bounds on it constrain their acceptable range.

The observables are measured by the Planck and other experiments for the comoving wavenumber
$k=k_* = a_{k_*}H_{k_*}$, with $a_{k_*}$ and $H_{k_*}$ denoting the scale factor and the Hubble scale at the moment of exit beyond the horizon of the mode with  comoving wavenumber $k_*$. One has an identity
\begin{equation}
   0 = \ln\left(\frac{k_*}{a_{k_*} H_{k_*}}\right)=\ln\left(\frac{a_{\rm end}}{a_{k_*}}\frac{a_{\rm re}}{a_{\rm end}}\frac{a_0}{a_{\rm re}}\frac{k_*}{a_0 H_{k_*}}\right),
    \label{eq:k=aH}
\end{equation}
where $a_{\rm end}$ and $a_{\rm re}$ are scale factors at the end of inflation and at the completion of reheating. The first three factors under the logarithm on the r.h.s.~of the above formula correspond to three periods of the universe evolution:
inflation, reheating and standard evolution after reheating, respectively.   
It is a consistency relation reflecting the fact that the observed perturbation left the (event) horizon at the scale factor $a_{k_*}$ and reentered the (particle) horizon when the scale factor had some known value (depending of the choice of the pivot scale $k_*$).
The number of e-folds $N_{k_*}=\ln\left(\frac{a_{\rm end}}{a_{k*}}\right)$ from the horizon exit of the mode $k_*$ to the end of inflation has already been given in eq.~\eqref{eq:Nk-integral} in terms of the inflaton  potential parameters and the value  of the inflaton $\phi_{\rm end}$ at the end of inflation, expressed by   
the CMB observables $n_s(k_*)$, $r(k_*)$ and $A_s(k_*)$. 
Regarding the reheating period, we first adopt the standard description by the Boltzmann equation of the energy density transfer to radiation, with the inflaton dissipation rate  $\Gamma$ as a free parameter. We trade $\Gamma$ for the reheating temperature $T_{\rm re}$.
It is then clear that for a given model of inflation the identity eq.~\eqref{eq:k=aH}  gives us the reheating temperature (or $\Gamma$) fixed in terms of  the observables
$A_s(k_*)$, $n_s(k_*)$ and $r(k_*)$.  
Conversely, some well motivated bounds on the reheating temperature, as we see later, can be translated into very severe tests of inflaton models by the CMB data.  Secondly,  in this paper we also investigate the dependence of the trajectories of fixed values of $T_\times$ in the $(n_s,  r)$ plane on some non-perturbative effects during the reheating period. 
Namely, we use numerical lattice simulations based on public package 
{{\tt ${\mathcal C}$osmo${\mathcal L}$attice}}
\cite{Figueroa:2020rrl,Figueroa:2021yhd} to investigate possible fragmentation  of the inflaton condensate during its oscillations around the minimum of the potential. In some cases such fragmentation may substantially change the reheating process so also predictions of a considered model. 
However, we start with 
the details of the perturbative part of the above outlined procedure,  anticipating  that in some models fragmentation plays no important role.
Inflaton fragmentation will be discussed in subsections \ref{subsec:n3n5} and \ref{subsec:n3/4n1/2} devoted to models for which fragmentation is important.

We define the reheating temperature $T_{\rm re}\equiv T_\times$ as the temperature 
of radiation at the moment when $\rho_\phi = \rho_R$, where $\rho_\phi$ and $\rho_R$ are the inflaton and radiation energy densities, respectively, that is, the time or the temperature at which the energy densities cross each other. 
The value of $T_\times$  enters into eq.~\eqref{eq:k=aH}  in the ratios $\left(\frac{a_{\rm \times}}{a_{\rm end}}\right)$  and $\left(\frac{a_0}{a_ \times}\right)$ (now $a_{\rm re}\equiv a_\times)$. The latter has the standard form:
\begin{equation}
\label{eq:T_rh_vs_N_rh}
    \begin{aligned}
        \frac{a_{\times}}{a_0}&=\left(\frac{43}{11g_{s*}}\right)^{1/3}\frac{T_0}{T_\times}\,,
    \end{aligned}    
    \end{equation}
where $T_0$ is the present temperature of the universe.   
The ratio $\left(\frac{a_\times}{ a_{\rm end}}\right)$ requires more attention.
Perturbative reheating is described by the following set of Boltzmann equations\footnote{We assume $\Gamma$ to remain constant during the entire reheating process.}:
\begin{eqnarray}
\dot{\rho_\phi} &=& - 3(1+w)H\rho_\phi - \Gamma \rho_\phi
\label{rho_phi_dot}
\,,\\
\dot{\rho_R} &=& - 4H\rho_R + \Gamma \rho_\phi
\label{rho_R_dot}
\,,\\
H^2 &=& \frac{\rho_\phi + \rho_R}{3\Mp^2}
\label{H_with_rho_R}
\,,
\end{eqnarray}
where dots denote derivatives with respect to the cosmic time. 
$\rho_\phi$ is the total energy density of the inflaton field which consists of two components: one from the oscillating homogeneous mode of the inflaton field and second from the relativistic inflaton particles (if present). The inflaton equation of state parameter $w$ depends on the shape of the inflaton potential and on the amount of inflaton particles. Thus, in general $w$ may change over time. 
For now we assume $w$ to be constant. Corrections resulting from production of inflaton particles due to fragmentation will be considered later in subsections \ref{subsec:n3n5} and \ref{subsec:n3/4n1/2}. 
In the leading order of the expansion in the inflaton field it is given by the following function of the potential exponent $n$: 
\begin{equation} 
\label{w}
w\approx \frac{n-1}{n+1}\;.
\end{equation}
It occurs that it is more convenient to analyze the above set of equations using the cosmic scale factor, $a$, instead of the cosmic time as the independent variable. 
We will apply also the usual approximation consisting in neglecting the radiation contribution to the total energy density during the reheating process, i.e.~$\rho_R\ll\rho_\phi$, resulting in the 
following approximate expression for the Hubble parameter
\begin{equation}
H^2 \approx \frac{\rho_\phi}{3\Mp^2}\,.
\label{H_approx}
\end{equation}
This way equations \eqref{rho_phi_dot} and \eqref{rho_R_dot} with $H$ given by \eqref{H_approx} may be rewritten in the following form:
\begin{eqnarray}
a\rho_\phi^\prime &=& -3(1+w)\rho_\phi - \sqrt{3}\Mp\Gamma\sqrt{\rho_\phi}
\label{rho_phi_prime}
\,,\\
a\rho_R^\prime &=& -4\rho_R + \sqrt{3}\Mp\Gamma\sqrt{\rho_\phi}
\label{rho_R_prime}
\,,
\end{eqnarray}
where primes denote derivatives with respect to the cosmic scale factor $a$.
The above set of equations may be solved analytically. The solution reads
\begin{eqnarray}
\rho_\phi
&=&
\rho_{\rm end}\left[(1+\gamma)\left(\frac{a}{a_{\rm end}}\right)^{-\frac32(1+w)}-\gamma\right]^2
\label{rho_phi_sol}
,\\
\rho_R
&=&
\rho_{\rm end}\,\frac{\Gamma}{H_{\rm end}}\left[\frac{2(1+\gamma)}{5-3w}\left(\left(\frac{a}{a_{\rm end}}\right)^{-\frac32(1+w)}-\left(\frac{a}{a_{\rm end}}\right)^{-4}\right)\right.
\nonumber\\
&&\qquad\qquad\quad
\left.-\frac{\gamma}{4}\left(1-\left(\frac{a}{a_{\rm end}}\right)^{-4}\right)\right]
,\qquad
\label{rho_R_sol}
\end{eqnarray}
where
\begin{equation}
\gamma=\frac{\Gamma}{3(1+w)H_{\rm end}}
\,,
\label{gamma_def}
\end{equation}
and we used obvious initial conditions at the beginning of reheating (identified with the end of inflation): $\rho_R(a_{\rm end})=0$, $\rho_\phi(a_{\rm end})=\rho_{\rm end}$. For a given model energy density at the end of inflation may be calculated in terms of the CMB observables. 
In the case of P-model with $n=1/2$ it reads
\begin{equation}
\rho_{\rm end }^{(n=1/2)} 
=
2\pi^2\Mp^4\,r\,A_s\,
\frac{\xi-1}{\xi+1}\,
\frac{\sqrt{4+\sqrt{r}\left(\xi^2-1\right)}-2}{\sqrt{4+\sqrt{r}\left(\xi^2-1\right)}+2}
\,.
\label{eq:rho_end_1/2}
\end{equation}
For P-models with other values of $n$ numerically obtained value of $\rho_{\rm end }$ will be used.

We define the end of reheating as the moment when the cosmic scale factor is equal $a_\times$ for which $\rho_\phi(a_\times)=\rho_R(a_\times)$.
It is possible to calculate $a_\times$ after neglecting terms proportional to $(a/a_{\rm end})^{-4}$ in \eqref{rho_R_sol}.
We obtain
\begin{equation}
\frac{a_\times}{a_{\rm end}}
\approx
\left[
\frac{\gamma\left(16+3(1+w)\sqrt{9-3w}\right)}{2(1+\gamma)(5-3w)}
\right]^{-\frac{2}{3(1+w)}}\,.
\label{a_times}
\end{equation}

The used above approximation is very good when two conditions are met. First: $w<\frac53$ which is always the case in the $\alpha$-attractor models considered in this paper. Second: $\gamma$ is not too large. We checked by solving numerically the Boltzmann equations \eqref{rho_phi_dot}--\eqref{H_with_rho_R} that accuracy of analytical results is very good if $\Gamma$ is not much bigger than about $0.1 H_{\rm end}$. The results for bigger values of the ratio $\Gamma/H_{\rm end}$ becomes gradually less precise but this causes no real problem for our analysis. For $\Gamma$ of the order of or bigger than $H_{\rm end}$ we are close to the instant reheating limit for which the whole analysis of the reheating process is no longer needed. The instant reheating limit comes down to the assumption that reheating is so rapid that no details of it matter.

Calculation of $T_\times$ with such simplifying assumption is straightforward.
Energy density at $a_\times$ equals $\rho_\times=\rho_\phi(a_\times)+\rho_R(a_\times)=2\rho_\phi(a_\times)$. Substituting $a_\times$ given by \eqref{a_times} into \eqref{rho_phi_sol} we obtain the following simple result
\begin{equation}
\rho_\times=\frac{\rho_{\rm end}}{2}\left[3(1+w)\gamma\,\frac{2+\sqrt{9-3w}}{5-3w}\right]^2
\,.
\label{rho_times}
\end{equation}
From the equality $\rho_\phi(a_\times)=\rho_R(a_\times)$ one may calculate the reheating temperature as a function of $\gamma$ (defined in eq.~\eqref{gamma_def})
\begin{equation}
T_\times^4
=
\frac{30}{g_*\pi^2}\,\frac{\rho_\times}{2}
=
\frac{30}{g_*\pi^2}\,\rho_{\rm end}
\left[\frac32(1+w)\gamma\,\frac{2+\sqrt{9-3w}}{5-3w}\right]^2
\,,
\end{equation}
which may be easily inverted:
\begin{equation}
{\gamma}
=
\frac{T_\times^2}{\sqrt{\rho_{\rm end}}}\,
\sqrt{\frac{2g_*}{15}}\frac{\pi(5-3w)}{3(1+w)(2+\sqrt{9-3w})}
\,.
\label{gamma}
\end{equation}
In calculations with fixed reheating temperature $T_\times$ one should replace $\gamma$ with the above function of $T_\times$.\footnote{Notice that our procedure avoids using the effective equation of state parameter $w_{\rm eff}$, often used in the literature.}

The result \eqref{a_times} is important for us because the ratio $a_\times/a_{\rm end}$, after substituting $\gamma$ given by \eqref{gamma}, may be directly used in the relation~\eqref{eq:k=aH} (with $a_{\rm re}\equiv a_\times$). 
Finally, using the Friedman equation and definitions of the parameters $r$ and $A_s$, one obtains the following expression for $H_k$
\begin{equation}
\label{eq:H_k_as_fun_of_inflaton_params}
    H_k = \frac{\pi}{\sqrt{2}} \Mp \sqrt{rA_s}\,.
\end{equation}
This completes our task of relating the values of the observables $n_s, r, A_s$ to a value of $T_\times$ in an $\alpha$-attractor inflation potential 
with a given exponent $n$. This is an important prediction  due to the following: Firstly, there are model independent bounds on the reheat temperature of the universe
\begin{equation} 
10\,\MeV\lesssim T_\times \lesssim 2\cdot10^{15}\,\GeV\,.
\label{eq:Tx-bounds}
\end{equation}
The lower bound is the BBN energy scale while the upper one 
follows from the fact that the consistency condition \eqref{eq:k=aH} with higher reheat temperatures has no physically meaningful solutions (it can be formally solved but only with negative number of e-folds during reheating).
They provide strong tests via the CMB data on the models of inflation. Secondly, that explicit link of the CMB data to the reheat temperature may have important particle physics implications.

Since at present only the  upper bound on $r$ is known experimentally,  it is interesting to illustrate the above  results in the plane ($n_s$, $r$) as a function $r(n_s)= f(n, A_s, n_s, T_\times)$, for 
several fixed values of the exponent $n$.  This is done in the next section.

In many models the allowed region in the ($n_s$, $r$) plane has only some partial overlap with the region corresponding to the often used condition $50\le N_k\le 60$. 
Clearly, the approach based on well motivated bounds on $T_\times$ may give significantly different results compared to an (essentially arbitrary) ansatz for $N_k$.

Our method described in this Section and used in the next Section to investigate considered models of inflation is based on the presented above analytical formulae. Some approximations and simplifications have been used to derive those formulae. For example, we used the slow roll approach to inflation; we neglected the fact that just after the end of inflation the inflaton potential is not exactly a monomial of the inflaton field, so for some time parameter $w$ is not exactly given by eq.~\eqref{w}; we neglected the contribution of $\rho_R$ to the Hubble parameter during the reheating process. 
With this in mind, one should ask how accurate the results obtained by this method are. In order to check this we compared some of our results with corresponding results obtained with more precise numerical calculations.\footnote{We thank Renata Kallosh and Andrei Linde for sharing some of their results and one of their computer codes.}
This comparison showed that our results are 
quite accurate. Typical differences of values of $n_s$ obtained both ways are not bigger than 0.0005.

%%%%%%%%%%%%%%%%%%%%%%%%%%%%%%%%%%%%%%%%%%%%%%%%%%%%%
\section{Results}
\label{sec:results}
%%%%%%%%%%%%%%%%%%%%%%%%%%%%%%%%%%%%%%%%%%%%%%%%%%%%%

Using the formalism  summarised in the previous section, with inflaton fragmentation taken into account when necessary, 
we present now the predictions for the CMB observables obtained for several values of the $n$ parameter, $n\in\{1/2, 3/4, 1, 2, 3, 5\}$ in the class P of $\alpha$-attractor 
models.\footnote{P type $\alpha$-attractor models with $n=1$ and $n=10$ were investigated in \cite{Ghoshal:2025ejg}.}

\begin{figure}[tb]
\includegraphics[scale=0.45]{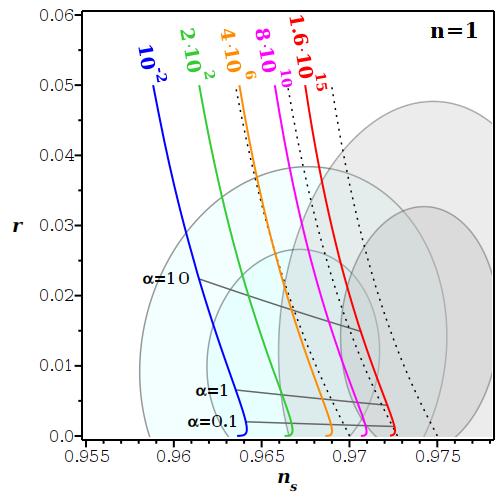}
\includegraphics[scale=0.45]{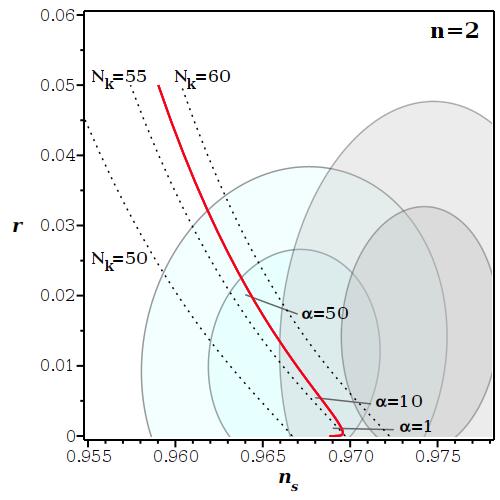}
\caption{Curves of constant reheating temperature $T_\times$ on the $(n_s, r)$ plane for $\alpha$-attractor P-models (color solid lines) with $n=1$ (left panel) and $n=2$ (right panel). Different colors correspond to different values of $T_\times$: $10^{-2}$ (blue), $2\cdot10^2$ (green), $4\cdot10^6$ (orange), $8\cdot10^{10}$ (magenta) and $1.6\cdot10^{15}$ (red) -- all in units of GeV. 
In the case of $n=2$ all five curves coincide. 
Dotted black curves show constant values of the number of e-folds, $N_k$, from the time when the pivot scale $k=a_0\times 0.05$\,Mpc$^{-1}$ crossed the Hubble horizon till the end of (slow roll) inflation. Curves for three values of $N_k$ are labeled on the right panel. Analogous curves for $N_k$ equal 50, 55 and 60 (from left to right) are shown also on all panels of figures \ref{fig:n=1n=2}, \ref{fig:n=3n=5} and \ref{fig:n=3/4n=1/2} but without labels. Some curves of constant values of the potential parameter $\alpha$ (see \eqref{eq:V}) are shown by gray lines. The shaded regions indicate $1\sigma$ and $2\sigma$ regions preferred by experiments. Light blue regions are for P-BK-D  \cite{BICEP:2021xfz} while gray regions are for  P-BK-D-ACT  \cite{AtacamaCosmologyTelescope:2025nti} combinations.
}
\label{fig:n=1n=2}
\end{figure}

We start with models for which inflaton fragmentation plays no important role, leaving more complicated cases to next subsections.  
The analysis is the simplest for $n=2$ i.e.~for models with inflaton potential which around its minimum may be approximated by a quartic function. Quartic potential results in two features following from the fact that in this case all three components of the energy density (inflaton oscillations, relativistic inflaton particles and relativistic SM particles) have the same $w=1/3$. 
First: the equation of state parameter $w$ during oscillations of the inflaton field is the same as during the RD era so the evolution of the scale factor does not depend on the reheating temperature $T_\times$. Second: the possible fragmentation of the inflaton condensate also has no impact on the expansion of the universe because production of relativistic inflaton particles does not change $w$. As a result for any value of $r$ only one value of $n_s$ is allowed, independently of the reheating temperature $T_\times$ (see right panel of Fig.~\ref{fig:n=1n=2}).

Analysis of models with quadratic inflaton potential, i.e.~with $n=1$, are also relatively simple. The reason is that there is no fragmentation during oscillations in a quadratic potential\footnote{We do not consider more complicated models in which strong coupling of the inflaton to other fields may lead to fragmentation also in quadratic potential \cite{Dufaux:2006ee,Antusch:2021aiw,Antusch:2022mqv}.}.
But, contrary to the $n=2$ case, there is $T_\times$ dependence of the results.
Now, during reheating $w=0$ so it is different from $w=1/3$ during RD period. Thus, the consistency condition \eqref{eq:k=aH} results in  nontrivial relation between $T_\times$ and the CMB parameters $n_s$ and $r$. Allowed curves on the $(n_s, r)$ plane are different for different $T_\times$ 
(see left panel of Fig.~\ref{fig:n=1n=2}). For a given value of $r$ the value of $n_s$ increases with $T_\times$ and changes by about 0.009 when $T_\times$ increases from the lower to the upper limit of the allowed range \eqref{eq:Tx-bounds}. 
For a given $T_\times$ the spectral index $n_s$ changes with $r$ and is maximal for $r\approx10^{-3}$.

A given model may be self-consistent only if it predicts a point on the ($n_s$, $r$) plane between the outermost colored curves because only then the consistency condition \eqref{eq:k=aH} and the model independent bound on the reheat temperature \eqref{eq:Tx-bounds} are fulfilled.\footnote{
The highest $T_\times$ presented in figures in this section equals $1.6\cdot10^{15}$\,GeV and is somewhat smaller than the model-independent upper bound \eqref{eq:Tx-bounds}. For $T_\times$ approaching the upper bound the length of the corresponding curve would shrink to a point quite close to the presented curve for $T_\times=1.6\cdot10^{15}$\,GeV. Observe that pairs of neighboring color curves in Figs.~\ref{fig:n=1n=2}, \ref{fig:n=3n=5} and \ref{fig:n=3/4n=1/2} correspond to temperatures differing by a factor of $2\cdot10^4$ while $T_\times=1.6\cdot10^{15}$\,GeV is only 20\% smaller than the upper limit in \eqref{eq:Tx-bounds}.}

For model building, of some interest are the values of the parameter $\alpha$  across the $(n_s, r)$ plane. They are indicated in the plots by gray lines. They change weakly with $n_s$, especially for small values of $r$. For fixed $T_\times$ they are monotonically growing functions of $r$. Value of $\alpha$ for a given pair of CMB parameters, $n_s$ and $r$, is for $n=2$ bigger then for $n=1$ (and even bigger for $n=3$ and $n=5$ -- see Fig.~\ref{fig:n=3n=5} in the next subsection). 
For example for $r\approx 0.02$ the values of $\alpha$ are: $\alpha={\cal{O}}(10)$ for $n=1$ and $\alpha\approx 50$ for $n=2$. Such behavior is mainly caused by the factor of $n^2$ in eq.~\eqref{eq:alpha}.

Fig.~\ref{fig:n=1n=2} shows that the model with $n=1$ can accommodate 
the P-BK-D data even at the $1\sigma$ level (for some range of $r$) in the full allowed range of reheating temperatures and P-BK-D-ACT data at $2\sigma$ level for temperatures in the range from about $2\cdot 10^2$\,GeV 
%(for $r$ approximately in the range $10^{-4}\div10^{-2}$) 
up to $\sim 10^{15}$\,GeV.  
For $n=2$, the temperature independent curve fits the P-BK-D combination in the whole range of $r$  and the P-BK-D-ACT combination for $r\lesssim 0.01$. The plots clearly show that the expected improvement in the upper bound  on $r$ at the level of $10^{-3}$ 
\cite{LiteBIRD:2022cnt}
will provide strong tests of those cases. 
One can see on the left panel of Fig.~\ref{fig:n=1n=2} that in the model with $n=1$ the range of values of the number of e-folds during inflation (after the pivot scale left the horizon), $N_k$, is substantially different from the usually assumed $50\div60$. The biggest possible $N_k$ depends on $r$ and may be only slightly bigger than 55. On the other hand, values of $N_k$ much smaller than 50 are allowed for low reheating temperature. $N_k$ may be as small as about $43$ for $T_\times=10$\,MeV.

Analysis of models with other values of $n$ is more complicated because in most cases the effects of fragmentation must be taken into account.
As a result of fragmentation of the inflaton condensate some of the energy stored in coherent inflaton oscillations is transferred to relativistic inflaton particles. The equation of state parameter of 
oscillating homogeneous mode of the inflaton field  
depends on the shape of the potential as $w=(n-1)/(n+1)$. Analogous parameter for relativistic particles is  $w=1/3$. The value of $w$ determines how fast the energy density of a given component decreases during the expansion of the universe. The larger $w$ the faster is dilution of energy. Thus,  for $n>2$  the energy of oscillations decreases faster than the energy of particles produced during fragmentation. As a result the energy of produced relativistic inflatons sooner or later dominate over the energy of oscillating condensate. The situation is opposite in models with $n<2$. Usually fragmentation is not fully complete so some energy stays in the form of coherent oscillations.  The parameter $w$ for such oscillations with $n<2$ is smaller than 1/3 so its energy decreases slower than the energy of produced relativistic inflaton particles. So,  such fragmentation results in only temporary increase of average $w$ (averaged over both contributions from the inflaton) which after some time returns to its initial value $w=(n-1)/(n+1)$.

As we see, effects of fragmentation are quite different in models with $n>2$ as compared to those with $n<2$. Thus, we will discuss them separately. First two models with $n=3, 5$ and then two models with fractional $n=1/2, 3/4$.
We consider models in which fragmentation is caused only by inflaton self-interactions, i.e.~we assume that no fields couple to the inflaton strongly enough to substantially modify the fragmentation process (see e.g.~\cite{Dufaux:2006ee,Antusch:2021aiw,Antusch:2022mqv}).

%%%%%%%%%%%%%%%%%%%%%%%%%%%%%%%%%%%%%%%%%%%%%%%%%%%%%%%%%%
\subsection{\boldmath$\alpha$-attractor P-models with $n>2$}
\label{subsec:n3n5}
%%%%%%%%%%%%%%%%%%%%%%%%%%%%%%%%%%%%%%%%%%%%%%%%%%%%%%%%%%

\begin{figure}[thbp]
\includegraphics[scale=0.45]{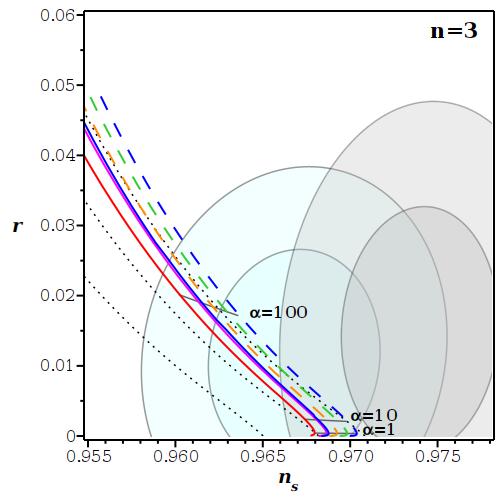}
\hfill
\includegraphics[scale=0.45]{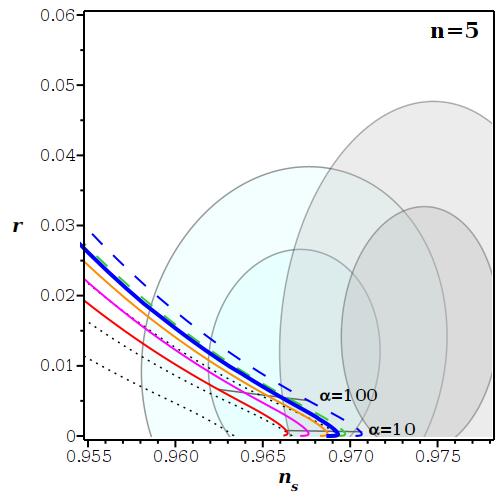}
\includegraphics[scale=0.45]{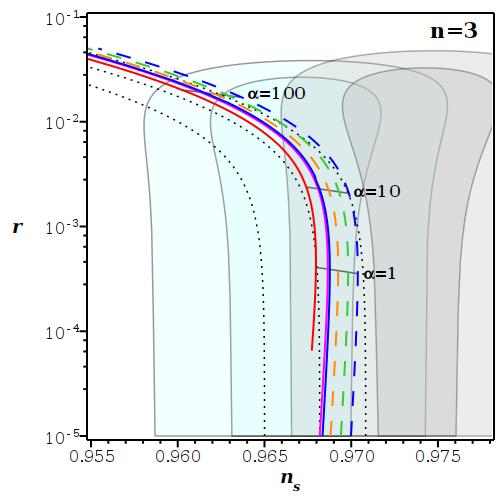}
\hfill
\includegraphics[scale=0.45]{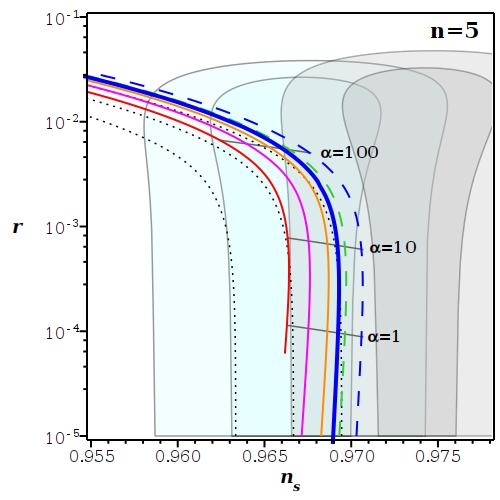}
\caption{Curves of constant reheating temperature $T_\times$ on the $(n_s, r)$ plane for $\alpha$-attractor P-models with $n=3$ (left panels) and $n=5$ (right panels). 
Notation as in Fig.~\ref{fig:n=1n=2} with two modifications. Dashed color lines correspond to fixed values of $T_\times$ which are obtained without effects of fragmentation taken into account. Thick blue lines show results with fragmentation included and for all values of $T_\times$ which in the approximation without fragmentation are to the right (dashed lines, bigger $n_s$). 
Same results are shown in linear (upper panels) and logarithmic (lower panels) vertical axes in order to show better both, relatively big and small, regions of $r$.
}
\label{fig:n=3n=5}
\end{figure}

First we discuss two P-type $\alpha$-attractor models with $n=3$ and $n=5$. 
Curves of fixed reheating temperatures $T_\times$ on the $(n_s, r)$ plane are shown in Fig.~\ref{fig:n=3n=5}. There are some differences with respect to the $n=1$ case. 
First: as for all models with $n>2$ the relation between the reheating temperature $T_\times$ and the spectral index $n_s$ is opposite to that for models with $n<2$.
Namely, for a given value of $r$ bigger values of $n_s$ correspond to lower $T_\times$. Second: the range of values of $n_s$ allowed for given $r$ with $T_\times$ changing in the full range \eqref{eq:Tx-bounds} is much smaller than for $n=1$ (and slightly bigger for $n=5$ than for $n=3$). Third: fragmentation of the inflaton condensate must be taken into account. 

The number of e-folds after which fragmentation starts strongly depends on the exponent $2n$ in the potential \eqref{eq:V} \cite{Garcia:2023dyf}. It is of order ${\cal{O}}(10)$ for $n=3$ and ${\cal{O}}(20)$ for $n=5$. 
And once it starts, the process proceeds relatively quickly, leading to the conversion of most of the inflaton field energy into relativistic inflaton particles. As a consequence, the parameter $w$ of the inflaton stays constant for quite long time and then changes quickly from the initial value of $(n-1)/(n+1)$ to 1/3. In our analysis we approximate this process by a step change of $w$
 at the number of e-folds after the beginning of reheating taken from  \cite{Garcia:2023dyf}.

The fragmentation (for $n>2$) influences those models for which duration of the reheating process calculated with only perturbative inflaton decay taken into account is bigger than the time at which fragmentation starts. Expansion of the universe after fragmentation is completed does not depend any longer on the reheating temperature because $w$ stays constant at the value 1/3 (until the end of the radiation domination era). As a result, there is one curve on the $(n_s, r)$ plane which corresponds to all reheating temperatures not bigger than such $T_\times$ for which the reheating completes just before fragmentation. This is a solid blue curve on each panel in Fig.~\ref{fig:n=3n=5}. The dashed lines to the right from it correspond to lower $T_\times$ obtained with fragmentation not taken into account. Those lines are plotted to illustrate the impact of fragmentation on the allowed region on the $(n_s, r)$ plane. Only the region between solid red and blue lines is allowed. It is much narrower for $n=3$ as compared with $n=5$ (range of $n_s$ for a given $r$ of about 0.001 and 0.003, respectively). There are two reasons for that. One: the range of $n_s$ allowed without fragmentation (between solid red and dashed blue curves) is narrower when $n$ is closer to $n=2$ (for which all curves coincide). Second: the effect of fragmentation is weaker for bigger $n$ because fragmentation starts later so it affect smaller range of $T_\times$. Fragmentation plays no role if $n\gtrsim7$ because for such $n$ the reheating process is completed before the onset of fragmentation even for the lowest considered $T_\times=10$\,MeV. For example, the results for P-model with $n=10$ presented in \cite{Ghoshal:2025ejg} are not altered by fragmentation.

Plots for $n=3$ and $n=5$ in Fig.~\ref{fig:n=3n=5} are shown in two versions, one with linear (upper panels) and second with logarithmic (lower panels) scale of the vertical axis. The logarithmic scale is used to show more clearly the results for small values of $r$. One can see on those plots that the red curves corresponding to $T_\times=1.6\cdot10^{15}$\,GeV end at $r$ slightly smaller than $10^{-4}$. The reason is that those end-points correspond to instant reheating. Solutions of the consistency condition \eqref{eq:k=aH} for even smaller $r$ would require negative number of e-folds during reheating which physically makes no sense. Curves describing other values of $T_\times$ also have such end-points but for very small $r$ (outside the ranges shown in our figures). For example magenta curves of $T_\times=8\cdot10^{10}$\,GeV end at $r={\cal{O}}(10^{-21})$.

Fig.~\ref{fig:n=3n=5} shows that for models with bigger $n$ experimental data lead to stronger upper bounds on $r$. Compatibility with the P-BK-D data at the $2\sigma$ level may be achieved only when $r\lesssim0.025$ ($r\lesssim0.018$) for $n=3$ ($n=5$). Compatibility with the combined P-BK-D-ACT data requires even smaller values of $r$  below $10^{-2}$. For sufficiently small $r$ both models, with $n=3$ and $n=5$, predict values of $n_s$ within $1\sigma$ region derived from P-BK-D  data and $2\sigma$ region of P-BK-D-ACT data for the full allowed range of the reheating temperatures. Compatibility with P-BK-D-ACT data is lost if $r$ is too small. For values of $r$ below about $10^{-12}$ the predicted value of $n_s$ becomes too small (such values of $r$ are outside the range shown in Fig.~\ref{fig:n=3n=5}).

Contrary to the case of $n=1$ now the range of values of the number of e-folds during inflation is quite narrow and those values are relatively big.  Namely, for both models with $n>2$ we obtain $N_k\gtrsim54$. The biggest possible values of $N_k$ depends on $n$: $N_k\lesssim60$ for $n=3$ and $N_k\lesssim63$ for $n=5$.

%%%%%%%%%%%%%%%%%%%%%%%%%%%%%%%%%%%%%%%%%%%%%%%%%%%%%%%%%%
\subsection{\boldmath$\alpha$-attractor P-models with $n<1$}
\label{subsec:n3/4n1/2}
%%%%%%%%%%%%%%%%%%%%%%%%%%%%%%%%%%%%%%%%%%%%%%%%%%%%%%%%%%

$\alpha$-attractor P-models with fractional values of $n$ are interesting 
for several reasons. One, discussed for example in \cite{Kallosh:2022feu}, is that such models may 
accommodate 
%lead to 
bigger values of the spectral index $n_s$ 
(as compared to models with $n\ge1$) 
which 
are supported %seems to be suggested 
by new experimental data \cite{BICEP2:2018kqh,BICEP:2021xfz}.
Inflaton condensate fragmentation is important for the analysis of such models. Moreover, fragmentation process for models with $n<2$, so also for $n<1$ considered in this subsection, proceeds in a different way than in case of models with $n>2$ (described in the previous subsection). Thus, we start with discussing this fragmentation in some detail.

One crucial difference between models with $n$ bigger and smaller than 2 is that the parameter $w$ describing oscillations in a potential which may be approximated by the monomial $|\phi|^{2n}$ (the absolute value of $\phi$ must be used in models with fractional $n$) 
%equals 
$w=(n-1)/(n+2)$,  so is smaller (bigger) than $1/3$ for $n<2$ ($n>2$).
The energy density of relativistic particles produced during fragmentation decreases faster (slower) than the energy density of the remaining condensate if $n<2$ ($n>2$). Thus, (contrary to the case with $n>2$) the remaining condensate plays important role in models with $n<2$ because sooner or later it starts to dominate over the produced inflaton particles.

\begin{figure}[tb]
\includegraphics[scale=0.45]{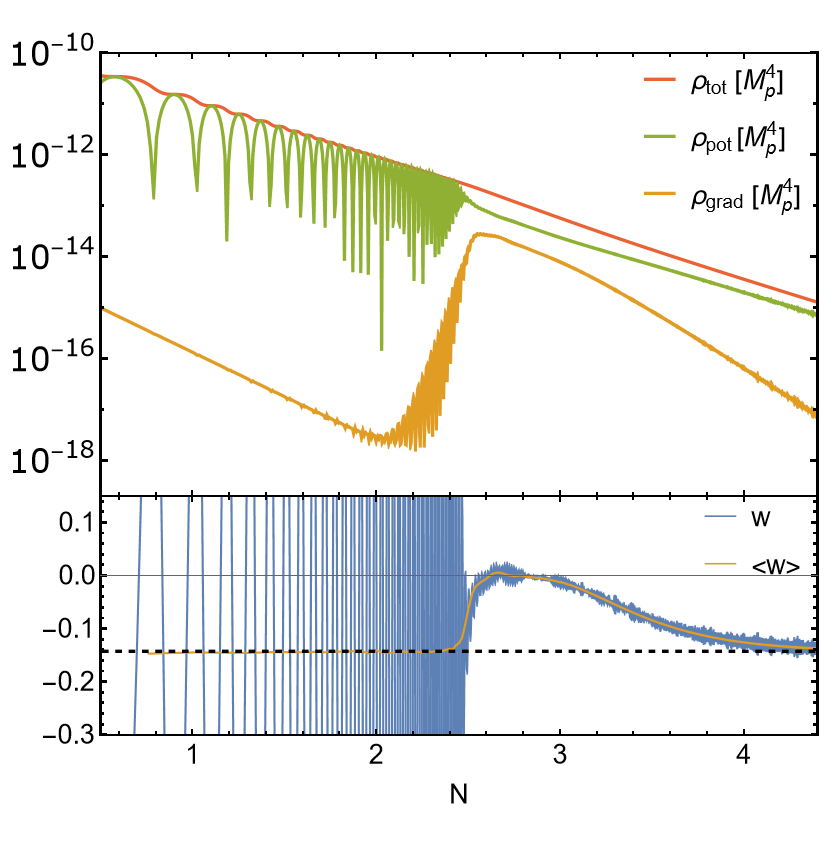}
\includegraphics[scale=0.45]{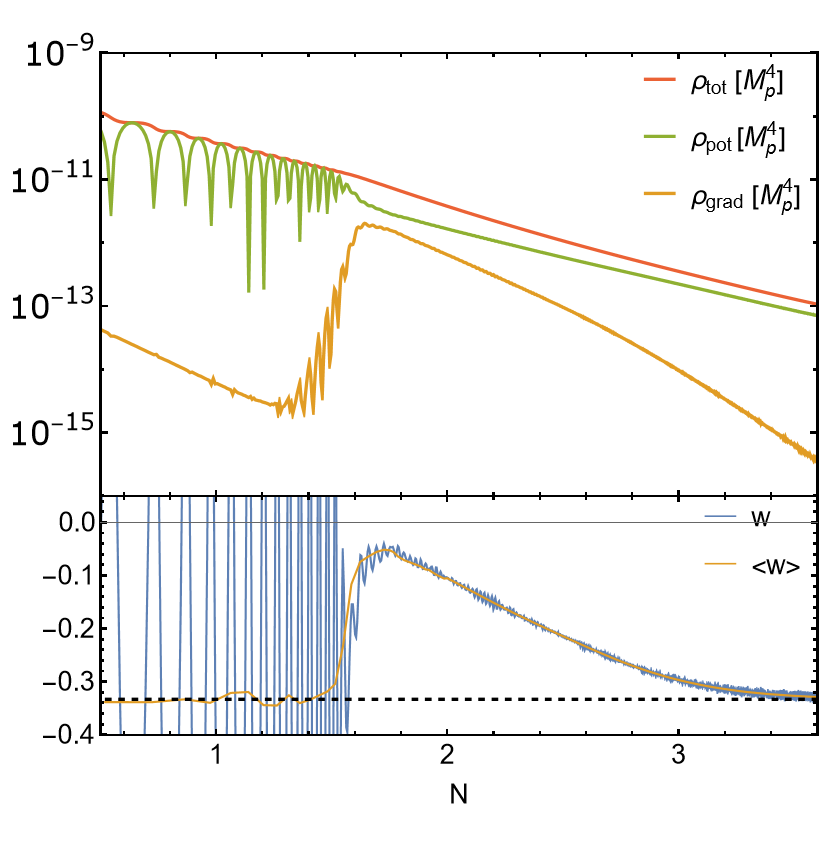}
\caption{Evolution of different components of the inflaton energy density (upper panels) and the resulting evolution of the inflaton equation of state parameter $w$ (lower panels) as functions of the number of e-folds from the beginning of the reheating process. Left panels are for model with $n=3/4$ while right panels are for model with $n=1/2$. In the upper panels shown are (from top down): total inflaton energy (red curve) and its potential (green) and gradient (orange) components. In the lower panels the blue curves show $w(N)$ and orange curves correspond to $w$ averaged over oscillation periods. Black dashed lines show value of $w$ corresponding to pure inflaton condensate i.e.:~-1/7 for $n=3/4$ and -1/3 for $n=1/2$.
In both cases parameters of the inflaton potential were chosen in such a way that without fragmentation they would give $T_\times=4\cdot10^{6}$\,GeV and $r=0.01$. 
The results were obtained by performing numerical simulations with the help of {{\tt ${\mathcal C}$osmo${\mathcal L}$attice}} program \cite{Figueroa:2020rrl,Figueroa:2021yhd} run on a $128^3$ lattice with $p_{\rm IR}/\omega_*=1$. 
}
\label{fig:lattice}
\end{figure}

The differences in fragmentation in different models are also reflected in different behavior of the $w$ parameter for the total inflaton energy (condensate plus particles). At early stage of reheating $w=(n-1)/(n+1)$. During fragmentation, due to the contribution from relativistic inflaton particles, the value of $w$ changes towards $1/3$. For $n>2$ this tendency remains also after the end of fragmentation and $w$ asymptotically approaches $1/3$. In models with $n<2$ the increase of $w$ is only temporary and after the end of fragmentation $w$ asymptotically returns to its initial value of $(n-1)/(n+2)$. Examples of such evolution obtained by numerical simulations using 
{{\tt ${\mathcal C}$osmo${\mathcal L}$attice}}
are shown in Fig.~\ref{fig:lattice}. 
Color curves in the upper panels in that figure show the evolution of total inflaton energy density (red curves) and also its potential (green curves) and gradient (orange curves) components (the kinetic component is not shown in order not to complicate the plots). The gradient component is to a very good approximation equal one half of the energy of produced relativistic inflaton particles (part of the kinetic component is the other half). Orange curves in Fig.~\ref{fig:lattice} illustrate the above discussed temporary increase of the importance of relativistic inflatons and the resulting temporary increase of the inflaton equation of state parameter $w$.

For models with $n<1$ discussed in this subsection a described above temporary increase of inflaton $w$ caused by fragmentation leads to faster decrease of inflaton energy density. As a consequence, energy density at the end of reheating (with a given perturbative decay rate) is smaller. Thus, for a given model taking into account the fragmentation process results in lower reheating temperature $T_\times$. 
For most of the temperature range \eqref{eq:Tx-bounds} considered in this work 
(for which the time scale of perturbative reheating is much longer than the time scale of fragmentation in models with $n<1$) 
the whole effect of fragmentation may be very well approximated by an appropriate change of the number of e-folds of the reheating process used in the consistency condition \eqref{eq:k=aH}, as compared to calculations with non-perturbative effects neglected.
A given value ot $T_\times$ determines the value of the energy density at the end of reheating. Due to fragmentation (in models with $n<2$) this energy density is achieved earlier than in the same model but with fragmentation neglected.\footnote{Of course, in order to fulfill the end of reheating condition $\rho_\phi=\rho_R$  earlier the inflaton decay rate $\Gamma$ must be bigger.}
The difference of the number of e-folds in both cases, $\Delta N$, may be approximated by
\begin{equation}
\Delta N
\approx
- \frac{\int (w(N)-w_0)\,\d N}{1+w_0}\,,
\label{eq:DeltaN}
\end{equation} 
where $w_0=(n-1)/(n+1)$ is the value of $w$ parameter of the inflaton before and well after fragmentation while $w(N)$ is the value of this parameter during fragmentation which changes due to changing contribution from produced inflaton particles.

In order to determine quantitatively the discussed above effects we performed numerical lattice simulations for several sets of parameters \eqref{eq:alpha}-\eqref{eq:Lambda4} typical for considered models with $n=1/2$ and $n=3/4$. 
The obtained results allowed us to find modification of the expansion of the universe and additional dilution of the inflaton energy density caused by fragmentation to be used 
in the consistency condition \eqref{eq:k=aH} in order to find the relation between CMB parameters and the reheating temperature. 
Numerical simulations allowed us to determine function $w(N)$ used in eq.~\eqref{eq:DeltaN} for each set of parameters. 
Resulting $\Delta N$ depends on values of the inflaton potential parameters, so also on $n_s$ and $r$, and for the considered models typically is ${\cal{O}}(1)$ or slightly smaller.
For the examples shown in Fig.~\ref{fig:lattice} we obtained 
$\Delta N\approx-0.35$ for $n=3/4$ and $\Delta N\approx-0.6$ for $n=1/2$.

\begin{figure}[tb]
\includegraphics[scale=0.45]{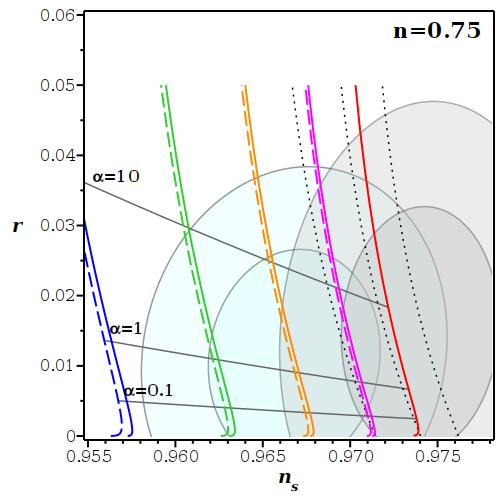}
\includegraphics[scale=0.45]{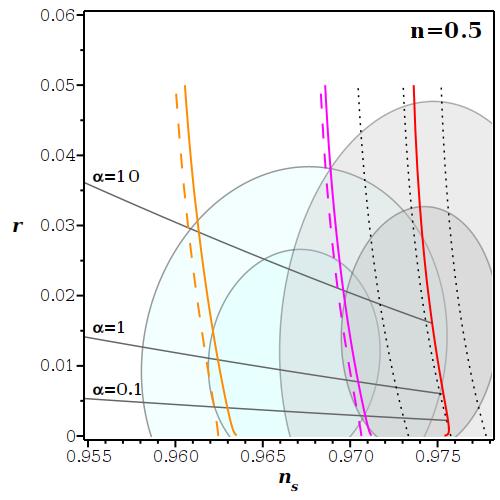}
\caption{Curves of constant reheating temperature $T_\times$ on the $(n_s, r)$ plane for $\alpha$-attractor P-models  with $n=3/4$ (left panel) and $n=1/2$ (right panel). Notation as in Figs.~\ref{fig:n=1n=2} and \ref{fig:n=3n=5} with one modification: additional dashed color lines show curves of constant $T_\times$ obtained with effects of fragmentation not taken into account.}
\label{fig:n=3/4n=1/2}
\end{figure}

The results of our analysis including fragmentation for two fractional values of $n$, namely $n=3/4$ and $n=1/2$, are presented in Fig.~\ref{fig:n=3/4n=1/2}. Curves of constant $T_\times$ with fragmentation taken into account are shown by solid color lines. They should be compared to dashed color curves obtained with fragmentation process neglected. The differences between solid and dashed lines for $n=3/4$ are smaller than for $n=1/2$. Such result could be anticipated because $n=3/4$ is closer to $n=1$ for which there is no fragmentation at all. Fragmentation is stronger in the case of $n=1/2$ but still not strong enough to produce very large corrections. For reheating temperature equal $4\cdot10^6$\,GeV inclusion of fragmentation leads to change of $n_s$ of order 0.001 (see orange lines on the right panel of Fig.~\ref{fig:n=3/4n=1/2}). Correction for $T_\times=8\cdot10^{10}$\,GeV (magenta lines) is even smaller.

Taking into account fragmentation effects in models with $n<1$ causes a shift of (some) lines of constant $T_\times$ towards larger values of $n_s$. The opposite tendency was observed in the previous subsection for models with $n>2$. The difference is due to the fact that $w_0=(n-1)/(n+1)$ is smaller (bigger) than 1/3 for $n<1$ ($n>2$) so inflaton particles produced during fragmentation cause the inflaton equation of state parameter $w$ to be bigger (smaller).

Comparing with the result for $n=1$ (left panel of Fig.~\ref{fig:n=1n=2}), one can see that maximal possible values of $n_s$, obtained for maximal reheating temperature and $r={\cal{O}}(10^{-3})$, are bigger. However differences are not very large. Such maximal $n_s$ is approximately equal $0.973$ for $n=1$ while it is around $0.974$ and $0.976$ for $n=3/4$ and $n=1/2$, respectively. Much bigger differences may be observed for lower $T_\times$. For model with $n=1$ the difference in predicted $n_s$ between the lowest and the highest $T_\times$ (distance between blue and red curves) is about $0.008$. Such difference grows quite quickly with decreasing $n$. For $n=3/4$ it is about $0.017$ while for $n=1/2$ it is almost $0.06$. This is the reason why on the right panel of Fig.~\ref{fig:n=3/4n=1/2} only three solid color curves are visible. Two remaining are far outside the shown range of $n_s$. For the lowest $T_x=10$\,MeV the predicted value of $n_s$ is close to 0.915 which is by far beyond the experimentally acceptable region.
Thus, in the case of small $n$ experimental data may be used to set lower bounds on $T_x$ which are stronger than the model independent requirement $T_\times>T_{\rm BBN}$. For example, for $n=1/2$ the $2\sigma$ limit from P-BK-D combination may be satisfied when $T_\times\gtrsim1.2\cdot10^5$\,GeV. The corresponding condition for the $n=3/4$ case is much weaker: $T_\times\gtrsim60$\,MeV.

Relatively wide ranges of values of $n_s$ which may be obtained in models with $n<1$ result in wide ranges of values of allowed $N_k$. The biggest $N_k$ is similar to that for $n=1$ i.e.~$N_k\lesssim56$. But the lowest possible values of $N_k$ are much smaller. The lower bounds on $N_k$ result from the experimental data (even smaller values could be accommodated in models with $n$ equal 3 and 5). We found the following bounds: $N_k\gtrsim36$ for $n=3/4$ and $N_k\gtrsim34$ for $n=1/2$.

There is a relatively narrow range of high $T_\times$ for which time scales of fragmentation and perturbative decay of inflaton are similar. Our procedure of combining lattice simulations with approximate analytical calculations described in Sec.~\ref{sec:results} can not be used in such situation. One should perform numerical simulations with both perturbative and non-perturbative effects taken into account simultaneously. Such calculations are beyond the scope of the present work. However, it is clear that corrections from fragmentation are in such cases smaller than for lower $T_\times$ because there is not enough time for fragmentation to fully develop.

For even higher $T_\times$ (for example for $T_\times=1.6\cdot10^{15}$\,GeV corresponding to red lines) there are no corrections at all because reheating completes before fragmentation could start.

%%%%%%%%%%%%%%%%%%%%%%%%%%%%%%%%%%%%%%%%%%%%%%%%%%%%%
\section{Conclusions}
%%%%%%%%%%%%%%%%%%%%%%%%%%%%%%%%%%%%%%%%%%%%%%%%%%%%%

In this paper there have been analysed constraints on the $\alpha$-attractor P-models, with different powers of the exponent $n$, following from the CMB data. The emphasis is on the role of the reheating temperature which has to satisfy model independent bounds, given from below by the BBN and from above by the theoretical self-consistency. For the reheating dynamics, both perturbative inflaton decays into radiation and non-perturbative  fragmentation of the inflaton condensate into relativistic inflaton particles are included. For certain values of $n$, the latter modifies the \textit{inflaton} equation of state parameter $w$  and for $n<1$ stronger bounds on the reheating temperature emerge than the model independent ones.   In an approach with the reheating temperature expressed directly in terms of the CMB observables, the bounds on it translate for each value of $r$ into narrow ranges of the spectral index $n_s$, dependent on the exponent  $n$. There are shown the results for $n=1/2, 3/4, 1, 2, 3, 5$, to illustrate the general patterns. It is shown that for all those values of $n$ both P-BK-D and P-BK-D-ACT combinations of experimental results can be accommodated at the $2\sigma$ level. However, the corresponding ranges of acceptable reheating temperatures are different in different cases.  
P-BK-D combination of the data is easily consistent with a broad range of parameters of the $\alpha$-attractor P-models, whereas the P-BK-D-ACT combination is more selective. Also one can clearly see that the expected from LiteBIRD sensitivity to $r\sim10^{-3}$ will be very important for further tests of the model.

\section*{Acknowledgments}
We thank Renata Kallosh and Andrei Linde for fruitful discussions  
and Anish Ghoshal and Paweł Kozów for collaboration on \cite{Ghoshal:2025ejg}.

\bibliographystyle{BiblioStyle}
\bibliography{P-models-CMB}

%%%%%%%%%%%%%%%%%%%%%%%%%%%%%%%%%%%%%%%%%%%%%%%%%%%%%%%%%%%%%%%
%%%%%%%%%%%%%%%%%%%%%%%%%%%%%%%%%%%%%%%%%%%%%%%%%%%%%%%%%%%%%%%
\end{document}